\documentclass{appolb}
\usepackage{graphicx}
\usepackage{bm}

\begin{document}
\title{Observation of the collective flow in proton-proton collisions}
\author{Piotr Bo\.zek \thanks{email: piotr.bozek@ifj.edu.pl}
\address{The H. Niewodnicza\'nski Institute of Nuclear Physics,
PL-31342 Krak\'ow, Poland} \address{
Institute of Physics, Rzesz\'ow University, 
PL-35959 Rzesz\'ow, Poland}}
\maketitle

\begin{abstract}
 The scenario of a  collective expansion of 
 matter created in proton-proton (p-p) collisions
 at the CERN Large Hadron Collider (LHC)   is discussed.
Assuming a small transverse size and a formation time of $0.1$fm/c  
of the source
we observe the build up  of a substantial transverse flow in 
relativistic 
hydrodynamic simulations.  In order to demonstrate the collectivity
 in p-p collisions 
we propose  to look at the
 multiplicity dependence of the elliptic flow coefficient.
 If  high multiplicity 
events originate from azimuthally asymmetric events containing two flux tubes, 
an observable signal above the statistical fluctuations  
in the measured elliptic
 flow could appear.
\end{abstract}



  Particles emitted from the fireball created in relativistic nuclear collisions
 exhibit both  thermal statistical emission and collective
 transverse flow. A strong indication of collectivity is seen in the  
azimuthally asymmetric flow. 
 The asymmetry in  momentum space is a consequence 
of the collective expansion of an azimuthally asymmetric source. Because 
the size of the system in p-p collisions is much smaller, the applicability 
of a strongly interacting fluid picture for the description of the fireball is 
less justified. Some signatures of statistical emission of particles are 
visible  in elementary collisions \cite{Becattini:2009sc}, but they 
could be explained by phase space dominance effects. An interesting similarity
 in  the spectra from p-p and Au-Au collisions has been noticed if
energy and momentum conservation effects are taken into account
 \cite{Chajecki:2009es}. 
Generally, from the observation of transverse momentum spectra 
alone it is difficult to demonstrate the presence of a collective flow,
 since qualitatively similar distributions could originate from statistical 
emission or some underlying individual particle production processes. 
The observation of the elliptic flow in p-p collisions would represent
a clear signature of collectivity. 
Two important questions arise when
 addressing this problem. First, for multiplicities expected in p-p 
collisions at LHC (or even more at the BNL Relativistic Heavy Ion Collider 
(RHIC) if the picture is applicable
 at these energies) statistical fluctuations in the distribution of particles
 would generate non-zero asymmetry. The second one is the unknown
 mechanism that could lead  to the 
asymmetry in the initial mini-fireball created in a p-p collision.

\begin{figure}
\includegraphics[width=.9\textwidth]{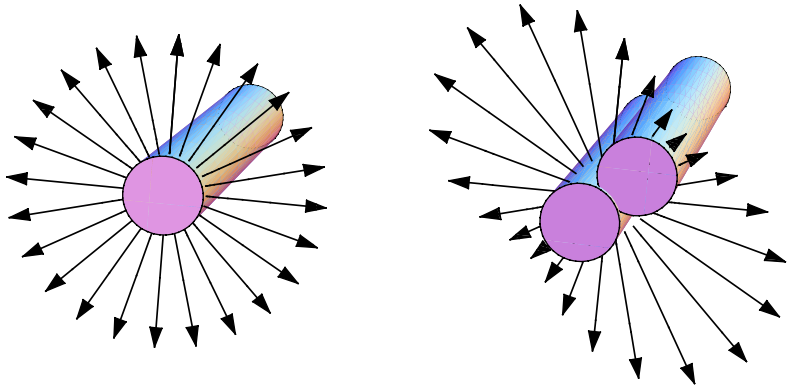}
\caption{Sketch of the one and two flux tubes configurations considered.
  On the 
left a single flux tube elongated in space-time rapidity generates 
azimuthally 
symmetric flow. On the right a configuration with two strings leads to an 
azimuthally  asymmetric flow in the transverse plane.}
\label{fig:ed}
\end{figure}

\begin{figure}
\begin{center}
\includegraphics[width=.45\textwidth]{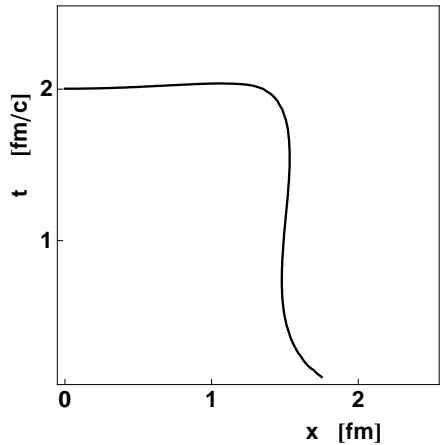}
\end{center}
\caption{Freeze-out hypersurface  in  the 
symmetric fireball corresponding to a p-p collision.}
\label{fig:freeze}
\end{figure}

The  small fireball in p-p collisions has been  treated in a 
similar way as in heavy-ion collisions 
\cite{d'Enterria:2010hd,Prasad:2009bx}, with an azimuthally asymmetric fireball
created in non-central collisions; 
another approach is proposed in Ref. \cite{Bautista:2009my}. All
 scenarios preadict a small
 elliptic flow in proton-proton collisions. In heavy-ion collisions
 many individual nucleons participate, and the resulting initial
 density can be approximated
as a continuous distribution, e.g. in the Glauber Model.
In p-p collisions there are only a few constituent partons 
\cite{Bialas:1977xp,Bjorken:1983}, and the description  based on the
 overlap of two densities in the colliding protons has no direct microscopic 
justification.

We follow a different approach.
According to the constituent quark model, one, two or 
rarely more independent sources are excited in a p-p collision.
The  source is assumed to produce particles 
in wide range of rapidities as in the string model or in 
bremsstrahlung emission. 
In a string picture of particle production particle are emitted 
in an azimuthally symmetric way \cite{Andersson:1983ia}. At LHC energies 
the string decay is fast  and the created matter is dense. A collective 
expansion stage could appear afterwards, 
 but one should not expect a significant geometrical asymmetry of the
 density in the transverse plane. We show that 
the  collective expansion of such dense matter created from the decay of a
single string-like object generates collective transverse flow, but without 
azimuthal asymmetry. The accumulated 
transverse flow  affects the observed spectra and the
 Hanburry Brown-Twiss (HBT) correlation radii
\cite{Humanic:2008zz}.
We propose  another signature of the collectivity in p-p 
collisions. It is the presence of the azimuthal asymmetry in the emission of 
particles in events containing two strings (flux tubes). 
The rapid decay of two flux tubes in the same event leads to
 an azimuthally asymmetric fireball where elliptic flow could be  generated
 in the expansion. The signal should be visible in high multiplicity events
(Fig. \ref{fig:ed}).

\begin{figure}
\begin{center}
\includegraphics[width=.45\textwidth]{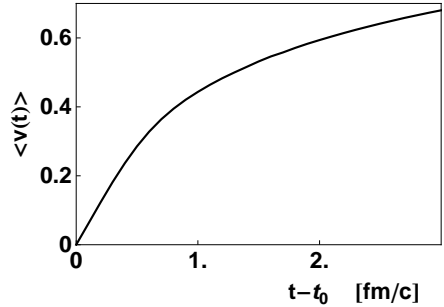}
\end{center}
\caption{Time dependence of the 
average transverse velocity in the   fireball.}
\label{fig:vsr}
\end{figure}

Let us first estimate how much of the transverse flow can be generated 
in the expansion of the small fireball corresponding to the system formed 
in p-p collisions at LHC energies. We use ideal fluid 
relativistic hydrodynamics
 to describe the dynamics of the fireball \cite{Kolb:2003dz}. 
We assume a Bjorken 
scaling flow in the longitudinal direction and a Gaussian profile of 
the energy density
\begin{equation}
\epsilon_{FT}(x,y)=\epsilon_0 \exp\left(-\frac{x^2+y^2}{2 \sigma^2}\right)
\label{eq:eone}
\end{equation}
in the transverse plane, $\sigma=0.5$fm.
The initial time for the expansion is $\tau_0=0.1$fm/c and the freeze-out
 temperature $140$MeV. The emission at freeze-out and the resonance decays are
 taken into account by the statistical emission code THERMINATOR 
\cite{Kisiel:2005hn} and a realistic equation of state of the plasma
 and hadronic phase is used \cite{Chojnacki:2007jc}.
The extrapolation
 of the charged particle multiplicity from  RHIC to LHC energies  
gives $\frac{dN}{d\eta}|_{\eta=0}\simeq 5$ \cite{Busza:2007ke}. 
To reproduce this number
 we adjust the value of $\epsilon_0=63$GeV/fm$^3$, which corresponds 
to a temperature of $450$MeV at the center of the fireball.

In Figure \ref{fig:freeze} is shown the freeze-out hypersurface. 
The lifetime of the fireball from the hydrodynamic
 evolution 
is of about $2$fm/c.
 A finite  time spent in the expansion of the system
is a necessary condition for the build up of the collective flow, since 
the transverse flow requires a minimal 
time of the order $\sigma/c_s$ to be generated, where $c_s\simeq\sqrt{3}$
 is the sound velocity.  In Fig. \ref{fig:vsr} is shown the average
transverse velocity in the system as function of time. The calculation 
shows that it is 
possible for the matter to acquire collective velocities of about $0.5$c 
during the short rapid expansion of the p-p collision fireball.
The presence of this collective velocity influences particle spectra, but
its extraction from this observable is not unambiguous.

\begin{figure}[htb]
\includegraphics[width=.65\textwidth]{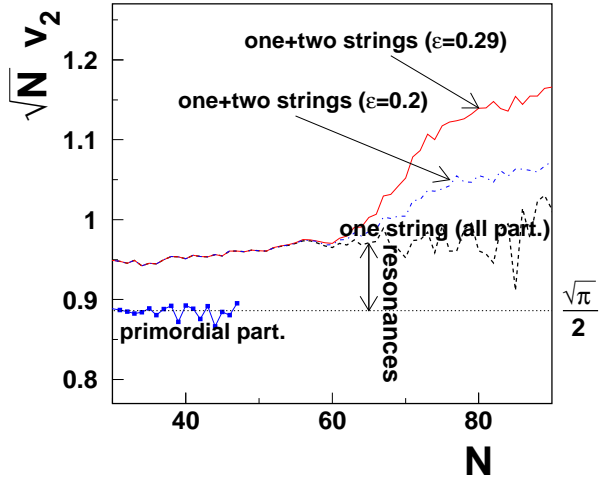}
\caption{Charged particle elliptic flow as function 
of the 
total charge multiplicity in the pseudorapidity interval  $|\eta|<4$. The 
solid and dash-doted lines represent the results for a mix of one and 
two-string events for two different initial separations of the two flux tubes
(two different initial eccentricities $e$).
The dashed line is the result when taking only one-string events, and the 
solid line with squares corresponds to primordial particles only  in
 one-string 
events. The results for primordial particles
 follow the prediction (\ref{eq:stat}) denoted 
by the dotted line.}
\label{fig:v2}
\end{figure}

Constituent quark models predict that in some of the collisions several
 constituents take part, in string models of particle 
production it corresponds to the excitation of two or more strings.
 The energy density distribution in an event with two strings is the sum of two individual profiles (\ref{eq:eone})
\begin{equation}
\epsilon_{FT}(x,y-d/2)+\epsilon_{FT}(x,y+d/2)
\label{eq:etwo}
\end{equation}
where $d$ is the separation between the two flux tubes. Depending on the 
distance $d=0.7$-$0.9$fm
 the energy distribution exhibits an eccentricity 
\begin{equation}
e=\frac{\langle y^2-x^2\rangle}{\langle x^2+y^2\rangle}
\end{equation}
of $0.2$-$0.29$. 
The proportion of events with two strings can be estimated in constituent 
quark models \cite{Bialas:1977xp,Bjorken:1983}. The result is reduced
 if constituent quarks are correlated as in the quark-diquark models
 \cite{Bialas:2006qf} and increased  if see quarks participate at 
higher energies. For the numerical estimate we take $20$\% of events 
with two strings. The final result depends weakly  on this number 
because the one and two-string events are separated in the total particle 
multiplicity (for large enough rapidity intervals and neglecting 
possible strong multiplicity fluctuations).
 The multiplicity in the events with two strings 
is on average almost double.

\begin{figure}[htb]
\includegraphics[width=.5\textwidth]{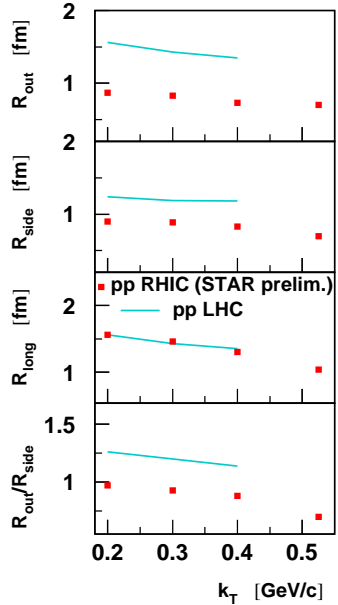}
\caption{Interferometry radii of pions as function of the transverse momentum 
of the pair from the hydrodynamic simulation of p-p collisions
 at LHC energies.
 The points represent preliminary data of the STAR Collaboration 
at RHIC $\sqrt{s}=200$GeV \cite{Chajecki:2005zw}.}
\label{fig:hbt}
\end{figure}

The elliptic flow coefficient of charged particles is calculated
in the pseudorapidity range $|\eta|<4$
\begin{equation}
v_2=\frac{1}{N}\sum_{i=1}^N \cos\left(2(\phi_i-\Psi \right) 
\end{equation}
in each event.
The {\it apparent} reaction 
 plane angle $\Psi$ 
is constructed in each event 
from the measured angular distribution to maximize $v_2$. 
This means that the elliptic flow is always positive due to the finite number
 of  particles. A simple model of azimuthally  symmetric, 
independent emission of
particles \cite{Broniowski:2007ft} gives
\begin{equation}
v_2=\frac{\sqrt{\pi}}{2\sqrt{N}} \simeq \frac{0.866}{\sqrt{N}}\ .
\label{eq:stat}
\end{equation}
This dependence on the total multiplicity $N$ is very different  from the one
 expected
 for a collective expansion of an asymmetric source. We plot the quantity 
$\sqrt{N} v_2$ in Fig. \ref{fig:v2}. For one string events (dashed line)
 we obtain 
$\sqrt{N}v_2\simeq 0.95$. It is more than expected for independent 
particles. This is due to correlations
 induced by resonance decays. The elliptic
 flow calculated from primordial
 particles (solid line with squares) follows 
the formula (\ref{eq:stat}). The  multiplicity of charged particles 
in one-string events is $45\pm 7$. For higher multiplicities we have
mainly two-string events. The hydrodynamic expansion of the 
asymmetric energy density (\ref{eq:etwo}) leads to a larger value
 of the elliptic flow (solid and dash-dotted lines in Fig. \ref{fig:v2}).
The elliptic flow saturates at high multiplicities instead of 
decreasing as $1/\sqrt{N}$. At very high multiplicities more 
than two strings could be excited and the elliptic flow would decrease
again.  Using the  reaction plane method, the trivial fluctuations from 
finite multiplicity (Eq. \ref{eq:stat}) can be easily subtracted.
Let us note that 
the contribution to the  elliptic flow from statistical fluctuations 
and resonances can be estimated and subtracted from the signal, also for
other experimental estimators of the elliptic flow coefficient 
using higher cummulants.

We calculate the HBT radii from a Gaussian fit to the correlation 
functions from the hydrodynamic model of the
 p-p collision \cite{Kisiel:2006is}. The finite life-time 
of the fireball leads to an increase of the interferometry radii 
(Fig. \ref{fig:hbt}) to about $1.5$fm. These values are larger than
 the measured
values of $R_{out}$ and $R_{side}$  
at RHIC for p-p collisions. Similar, larger
 values of the radii have been predicted 
 in a rescattering model \cite{Humanic:2008zz}. We confirm this observation 
in the hydrodynamic model of the  expansion of the fireball.  For  
pairs with the
 highest momentum a Gaussian fit cannot be obtained.

In summary, we use a hydrodynamic model of the dynamics of the small fireball
 created in p-p collisions at LHC energies. We conclude that during the
 lifetime of $2$fm/c significant transverse flow builds up. This has as
 a consequence the increase of  HBT radii. We suggest to identify the
elliptic flow in events with non-zero initial eccentricity. Such events could
occur due to the excitation of two flux tubes. 
The elliptic flow of charged particles measured in a wide pseudorapidity range
 would show a departure from the one resulting from statistical fluctuations
 only. This would happen at higher multiplicities where two-string events 
dominate.

After the competition of this paper, Ref. \cite{CasalderreySolana:2009uk}
appeared. The authors discuss a similar idea for the eccentricity
 fluctuations in the 
initial state, 
but with possibly many 
small string-like objects excited in the interaction region of a 
proton-proton collision.

\noindent
{\bf Acknowledgments}\\
Supported by 
Polish Ministry of Science and Higher Education under
grant N202~034~32/0918

\bibliography{../hydr}

\end{document}